\documentclass[aps,prb,preprint,nopacs,superscriptaddress]{revtex4}

\usepackage{graphicx}
\usepackage{verbatim}
\usepackage{mathrsfs}
\usepackage{array}
\usepackage{epstopdf}
\newcolumntype{P}[1]{>{\centering\arraybackslash}p{#1}}

\usepackage{amsmath,amsfonts,amssymb}
\usepackage{graphicx}

\def\3{2.8in}    
\def\2{2.5in}
\def\4{3.0in}

\def \beq {\begin{equation}}
\def \eeq {\end{equation}}

\begin{document}

\title{Discovery of Weyl fermion semimetals and\\ topological Fermi arc states}

\author{M. Zahid Hasan\footnote{Email: mzhasan@princeton.edu}}\affiliation {Laboratory for Topological Quantum Matter and Spectroscopy (B7), Department of Physics, Princeton University, Princeton, New Jersey 08544, USA} \affiliation{Lawrence Berkeley National Laboratory, Berkeley, CA 94720, USA}
\author{Su-Yang Xu}\affiliation {Laboratory for Topological Quantum Matter and Spectroscopy (B7), Department of Physics, Princeton University, Princeton, New Jersey 08544, USA}
\author{Ilya Belopolski}\affiliation {Laboratory for Topological Quantum Matter and Spectroscopy (B7), Department of Physics, Princeton University, Princeton, New Jersey 08544, USA}
\author{Shin-Ming Huang}
\affiliation{Department of Physics, National Sun Yat-Sen University, Kaohsiung 80424, Taiwan}
\affiliation {Laboratory for Topological Quantum Matter and Spectroscopy (B7), Department of Physics, Princeton University, Princeton, New Jersey 08544, USA}

\begin{abstract}

Weyl semimetals are conductors whose low-energy bulk excitations are Weyl fermions, whereas their surfaces possess metallic Fermi arc surface states. These Fermi arc surface states are protected by a topological invariant associated with the bulk electronic wavefunctions of the material. Recently, it has been shown that the TaAs and NbAs classes of materials harbor such a state of topological matter. We review the basic phenomena and experimental history of the discovery of the first Weyl semimetals, starting with the observation of topological Fermi arcs and Weyl nodes in TaAs and NbAs by angle and spin-resolved surface and bulk sensitive photoemission spectroscopy and continuing through magnetotransport measurements reporting the Adler-Bell-Jackiw chiral anomaly. We hope that this article provides a useful introduction to the theory of Weyl semimetals, a summary of recent experimental discoveries, and a guideline to future directions.


\end{abstract}

\pacs{Keywords: Weyl fermion, Semimetal, Topological Insulator, Chern number, Fermi surface}

\date{\today}

\maketitle

\section{\textbf{INTRODUCTION}}

The rich correspondence between high-energy particle physics and low-energy condensed matter physics has been a source of insights throughout the history of modern physics \cite{Anderson, Wilczek}. It has led to important breakthroughs in many aspects of fundamental physics, such as the Planck constant and blackbody radiation, the Pauli exclusion principle and magnetism, as well as the Anderson-Higgs mechanism and superconductivity, which in turn helped develop materials with useful applications. In the past decade, the discovery of massless Dirac fermions in graphene and on the surface of topological insulators has taken center stage in condensed matter science and has led to novel considerations of the Dirac equation in crystals\cite{Graphene,Hasan2010,Qi2011,Weyl_Vafek,NSrev,Volovik2003,TI_book_2014_2,Hasan long, HasanMooreARCMP2011}. The Dirac equation, proposed in 1928 by Paul Dirac, represents a foundational unification of quantum mechanics and special relativity in describing the nature of the electron. Its solutions suggest three distinct forms of relativistic particles: the Dirac, Majorana, and Weyl fermions. Only one year later, in 1929, Hermann Weyl pointed out that the Dirac equation without mass naturally gives rise to a simpler Weyl equation,whose solutions are associated with massless fermions of definite chirality, particles known as Weyl fermions \cite{Weyl}. Weyl’s equation was intended as a model of elementary particles, but in nearly 86 years, no candidate Weyl fermions have ever been established in high-energy experiments. Neutrinos were once thought to be such particles but were later found to possess a small mass. Recently, emergent, quasiparticle analogs of Weyl fermions have been discovered in certain electronic materials that exhibit strong spin–orbit coupling and topological behavior. Just as Dirac fermions emerge as signatures of topological insulators \cite{Hasan2010,Qi2011}, in certain types of semimetals, such as tantalum or niobium arsenides, electrons behave like Weyl fermions.

In 1937, the physicist Conyers Herring considered the conditions under which electronic states in solids have the same energy and crystal momentum even in the absence of any particular symmetry \cite{herring_accidental_1937}. Near such accidental band touching points, the low-energy excitations are described by the Weyl equation \cite{herring_accidental_1937,abrikosov_properties_1971,nielsen1983adler}. In recent times, these touching points have been studied theoretically in the context of topological materials and are referred to as Weyl points, and the quasiparticles near them are the emergent Weyl fermions \cite{ran, Murakami2007, Wan2011, Balents_viewpoint, Ashvin, Burkov2011, Burkov_NL, Huang2015, HgCrSe, TlBiSe, Xu_TPT, Weng2015}. In these crystals, the quantum mechanical wave function of an electron state acquires a geometric or Berry phase when tracing out a closed loop in momentum space. This Berry phase is identical to that acquired by an electron tracing out a closed loop in real space in the presence of a magnetic monopole. In the same way that magnetic monopoles correspond to sources or sinks of magnetic flux, a Weyl semimetal hosts momentum space monopoles that correspond to sources or sinks of Berry curvature. These Berry curvature monopoles are precisely the Weyl points of a Weyl semimetal. Furthermore, the chirality of the Weyl fermion corresponds to the chiral charge of the Weyl point. In a Weyl semimetal, the chirality associated with each Weyl node can be understood as a topologically protected charge, hence broadening the classification of topological phases of matter beyond insulators. Remarkably, Weyl nodes are extremely robust against imperfections in the host crystal and are protected by the crystal's inherent translational invariance. The real-space Weyl points are associated with chiral fermions, and in momentum space they behave like magnetic monopoles. The fact that Weyl nodes are related to magnetic monopoles suggests that they will be found in topological materials that are in the vicinity of a topological phase transition. Like a topological insulator, a Weyl semimetal hosts topological surface states arising from a bulk topological invariant. However, while the surface states of a topological insulator have a Fermi surface that consists of closed curves in momentum space, a Weyl semimetal hosts an exotic, anomalous surface-state band structure containing topological Fermi arcs, which form open curves that terminate on bulk Weyl points \cite{Wan2011,Balents_viewpoint,Ashvin}. Theory has suggested that in systems where inversion or time-reversal symmetry is broken, a topological insulator phase naturally allows a phase transition to a Weyl semimetal phase. Building on these ideas, many researchers, including the Princeton University group, used \textit{ab initio} calculations to predict candidate materials and perform angle-resolved photoemission spectroscopy (ARPES) to detect bulk Weyl points and surface Fermi arcs in TaAs and its cousins\cite{Weng2015, Huang2015, Hasan_Na3Bi, Hasan_TaAs, NbAs_Hasan, TaP_Hasan, TaAs_Ding_2, TaAs_Ding}. ARPES is an ideal tool for studying such a topological material, as known from the extensive body of work on topological insulators \cite{Hasan2010,Qi2011,HasanMooreARCMP2011}. The ARPES technique involves shining light on a material and measuring the energy, momentum, and spin of the emitted photoelectrons, both from the surface and the bulk. This allows for the direct observation of both bulk Weyl points and Fermi arc surface states. Weyl semimetals further give rise to fascinating phenomena in transport, including a chiral anomaly in the presence of parallel electric and magnetic fields, a novel anomalous Hall response, and exotic surface-state quantum oscillations. Even more exotic effects may arise in the presence of superconductivity, where Weyl semimetals may give rise to quasiparticles exhibiting non-Abelian statistics, potentially providing a platform to realize novel effects in spintronics or a new type of topological qubit \cite{Weyl-SC-7, Weyl-SC-1, Weyl-SC-5}.

In this review, we survey the experimental discovery of the first Weyl semimetal in TaAs. We first provide some key elements of the theory of Weyl semimetals. Then, we offer a history and some intuition for Weyl semimetals from the point of view of a search problem. Next, we review the key theoretical and experimental works of the discovery of the first Weyl semimetal in TaAs by ARPES. We also discuss the discovery of Weyl semimetals in the other compounds of the TaAs family, namely NbAs, TaP, and NbP. As an application of the ARPES results, we discuss observation of the chiral anomaly in TaAs by transport. We further mention briefly, but in no way attempt to discuss exhaustively, a wide range of closely related topics, including: inversion-breaking Weyl semimetals beyond the TaAs class, topological line node semimetals, strongly Lorentz violating Type II Weyl semimetals, magnetic Weyl semimetals, and Weyl superconductors. These additional directions will no doubt continue to enrich a fascinating and rapidly developing field of research.

\bigskip
\bigskip
\bigskip

\section{\textbf{Key Concepts}}

Before reviewing the experimental discovery of Weyl fermion semimetals, we present a few key concepts useful to understand what follows. We do not by any means attempt to provide a complete review of the current theoretical understanding of Weyl semimetals.\\

We present an explicit theoretical model for a Weyl semimetal. We consider a two-band Bloch Hamiltonian in three dimensions,
\begin{gather}
h(k) = a(k) \sigma_x + b(k) \sigma_y + c(k) \sigma_z \nonumber \\
a(k) = t (\cos (k_x) + \cos (k_y) + \cos (k_z) - \cos(k_0) - 2), \\
b(k) = t \sin (k_y)\textrm{, \ \ \ \ }c(k) = t \sin (k_z) \nonumber
\end{gather}
The lattice constant is set to unity. Here, the $\sigma$ are the Pauli matrices, $k = (k_x, k_y, k_z)$ is the crystal momentum, $t$ is a hopping parameter and $h(k)$ is the $2 \times 2$ Bloch Hamiltonian matrix. We can see that at $k_W^\pm=(\pm{k_0},0,0)$, all three of the functions $a(k)$, $b(k)$ and $c(k)$ simultaneously vanish. At these special $k_W^\pm$ the two bands are degenerate.

We can expand the Hamiltonian in the vicinity of the $k_W^\pm$ to derive a low-energy effective model. We note that $a(k) \approx (k - k_W^\pm) \cdot \nabla a(k_W^\pm)$ and similar for $b(k), c(k)$. We define $p^\pm \equiv k - k_W^\pm$ and we find that the low-energy model gives a cone-shaped dispersion,
\begin{gather}
h^\pm(p) = v_x p^\pm_x \sigma_x + v_y p^\pm_y \sigma_y + v_z p^\pm_z \sigma_z \\
v_x = - t \sin (\pm k_0) \textrm{, \ \ \ \ } v_y = v_z = t \nonumber
\end{gather}

Our result is tantamount to the Weyl Hamiltonian of particle physics \cite{Weyl},
\begin{equation}
H_W(p) = p \cdot \sigma
\end{equation}

However, we note that unlike $H_W(p)$, our effective theory $h^\pm(p)$ shows a Weyl fermion which has different Fermi velocities in different momentum directions, and so is not isotropic. This behavior is quite reasonable because Lorentz invariance is not required in low-energy effective theories in crystals. Next, we consider the effect of adding small, arbitrary perturbations to $h(k)$, taking $a'(k) = a(k) + \Delta a(k)$ and similar for $b(k), c(k)$. Will such perturbations gap out the Weyl cones? To lowest order, we find $a'(k) \sim (k - k_W^\pm) \cdot \nabla a'(k_W^\pm) + \Delta a(k_W^\pm) = (k - (k_W^\pm + \Delta k^\pm)) \cdot \nabla a'(k_W^\pm)$, where $\Delta k^\pm$ in general will depend on all three of the perturbing functions $\Delta a(k)$, $\Delta b(k)$ and $\Delta c(k)$. We see that the low-energy Hamiltonian remains gapless, but that the Weyl points move around in momentum space as we perturb the system.

The local stability of the Weyl points is related to the Chern number, a topological invariant well-known from the integer quantum Hall effect. The Chern number $\chi$ is defined as,
\begin{gather}
\chi = \frac{1}{2\pi} \iint dk_x dk_y \hat{z} \cdot \Omega \textrm{, \ \ \ \ } \Omega = \nabla_k \times A \textrm{, \ \ \ \ } A = - i \langle k, n | \nabla_k | k, n \rangle
\end{gather}
Here, $A$ is the Berry connection and $\Omega$ is the Berry curvature. In the case of the integer quantum Hall effect, the system is two-dimensional, so we integrate over the entire Brillouin zone. To find the Hall conductivity of a quantum Hall state, we further sum the result over all occupied bands $n$. In the case of a Weyl semimetal, the system is three-dimensional, so we must choose some closed two-dimensional manifold within the bulk Brillouin zone and calculate the Chern number on that manifold. If we choose a small spherical manifold enclosing the Weyl point, we find that the Weyl point is associated with a Chern number $\chi = \pm1$. We refer to this $\chi$ as the chiral charge of the Weyl point. The chiral charge measures the Berry flux through the spherical manifold, analogous to Gauss's law in classical electrodynamics. In the same way as for an electric charge, the chiral charge is quantized and the Chern number on any manifold depends only on the enclosed chiral charge.

It is clear that Weyl points arise generically in two-band models. In particular, the underlying mathematics does not care about the basis of the Hamiltonian. Typically, a Weyl semimetal refers to a normal electron system, but the basis can also be a Bogoliubov spinor, describing a Weyl superconductor, or it may consist of bosonic particles, describing, for instance, a photonic Weyl semimetal. At the same, it is important to note that most normal electron crystals have both time reversal and inversion symmetry, which requires all bands to be everywhere doubly-degenerate. In such a system, a band crossing corresponds to a four-fold degeneracy, requiring a description by a four-band model. However, in a four-band model, there are too many Hamiltonian matrix elements that must be simultaneously set to zero, so band crossings do not arise generically. In this way, the experimental study of Weyl semimetals has long been held back because nature prefers to maintain both time reversal and inversion symmetries.

If Weyl points are topologically stable, we might ask how they can be created or destroyed. While an individual Weyl point cannot be gapped out within band theory by small perturbations, a large perturbation can cause Weyl points of opposite chirality to annihilate each other, leaving the system gapped. If we imagine slowly applying a large perturbation to $h(k)$, we will find that the Weyl points move around in momentum space until the system arrives at a critical point where the Weyl points sit on top of each other. A further perturbation can then cause the system to gap out. Conversely, Weyl points can only be created through such a critical point, in sets of equal and opposite chiral charge. An equivalent statement is that the net chiral charge in the entire Brillouin zone is always zero. It is also possible to gap a Weyl point by going beyond band theory. If $h(k)$ is taken to describe a normal electron system, then superconductivity may gap the Weyl points by breaking $U(1)$ charge conservation symmetry. A charge density wave or disorder may also allow scattering directly between the $k_W^\pm$, gapping the Weyl cones by breaking translation symmetry. In this sense, Weyl points are protected by charge and translation symmetries. By contrast, the Dirac points in graphene\cite{Graphene}, topological insulators\cite{Hasan2010,Qi2011} and Dirac semimetals\cite{Hasan_Na3Bi} typically require not only the symmetries implied within band theory, but further rely on time reversal symmetry, inversion symmetry or other crystal space group symmetries. We see that Weyl points are uniquely robust in that they are topologically stable within band theory without any additional symmetries.

The integer quantum Hall effect is associated with gapless chiral edge modes guaranteed by the Chern number. What boundary states are guaranteed by Chern numbers in Weyl semimetals? We consider two-dimensional slices of the bulk Brillouin zone, as shown in Fig.~\ref{Weyl_Cartoon}\textbf{c-e}. Any slice not containing the Weyl points is gapped, and we can calculate a Chern number for that slice. In this way, we can view the three-dimensional band structure as a set of two-dimensional slices with a tuning parameter, $k_x$. As we scan $k_x$, the bulk band gap closes and reopens and we tune the system through a topological phase transition, changing the Chern number. The critical slices $k_x = \pm k_0$ are the slices containing a Weyl point. We can similarly partition the surface states of a Weyl semimetal into one-dimensional edges, as shown in the surface Brillouin zone square in Fig.~\ref{Weyl_Cartoon}\textbf{d}. Edges associated with a non-zero Chern number in the bulk will host gapless chiral edge modes, Fig.~\ref{Weyl_Cartoon}\textbf{e}, while edges with a zero Chern number are gapped, Fig.~\ref{Weyl_Cartoon}\textbf{c}. We can see that these one-dimensional edge states then assemble into a sheet of surface states which terminate at the surface projections of the Weyl points, forming a topological Fermi arc surface state. On a constant-energy cut of the surface band structure, the Fermi arc forms an open, disjoint curve. Much like the Dirac cone surface state of a three-dimensional $\mathbb{Z}_2$ topological insulator, the Fermi arc of a Weyl semimetal is anomalous in the sense that it cannot arise in any isolated two-dimensional system, but only on the two-dimensional boundary of a three-dimensional bulk. However, the Fermi arc arguably provides the most dramatic example to date of an anomalous band structure, because unlike the Dirac cone surface state or any traditional two-dimensional band structure, the constant-energy contours do not even form closed curves.

\bigskip
\bigskip
\bigskip
\section{\textbf{SEARCH FOR MATERIALS}}

The paramount experimental challenge to the discovery of the first Weyl semimetal was finding suitable material candidates. It was necessary to find compounds available in high-quality single crystals with Weyl fermions and Fermi arcs accessible under reasonable measurement conditions. In the five years or so preceding the discovery of TaAs, roughly from 2011 to 2015, the fundamental theory of Weyl semimetals was essentially available and powerful ARPES systems were ready to tackle the problem. The missing link was the lack of material candidates. In this section, we discuss the particular insights relating to material search that bridged theory to experiment and directly opened the experimental study of Weyl semimetals. We also highlight some open problems along these directions.

It was perhaps a historical accident that the community initially focused on time-reversal breaking Weyl semimetals. After Murakami's work showing a connection between Weyl semimetals and topological insulators\cite{Murakami2007}, Wan \textit{et al.} proposed in 2011 the first material candidate for a Weyl semimetal in a family of magnetic pyrochlore iridates, $R_2$Ir$_2$O$_7$, where $R$ is a rare-earth element \cite{Wan2011}. In this crystal, the Ir atoms form a sublattice of corner-sharing tetrahedra and the authors argued that the material prefers an all-in/all-out configuration of magnetic moments on the tetrahedra, breaking time-reversal symmetry \cite{Wan2011}. Next, theoretical analysis showed that, by increasing the on-site Coulomb interaction strength, the system exhibits a transition from a magnetic metal to a Mott insulator. In between, there is an intermediate phase, which was found to be a Weyl semimetal. Wan \textit{et al.} also made an explicit connection between Weyl semimetals and the integer quantum Hall (IQH) state. Specifically, they proposed the idea of calculating Chern numbers on two-dimensional manifolds in the three-dimensional Brillouin zone of a bulk material. In a Weyl semimetal, there would be manifolds with nonzero Chern numbers. As in the IQH effect, the one-dimensional edge of the two-dimensional manifold would protect chiral edge states and these would assemble together to form Fermi arc surface states on the two-dimensional surface of a Weyl semimetal. The specific material proposal and theoretical advances spurred considerable interest in Weyl semimetals. However, attempts to realize a Weyl semimetal in $R_2$Ir$_2$O$_7$ in experiment were met with significant challenges. The all-in/all-out magnetic order is under debate in experiments \cite{Iridate_Mag_1, Iridate_Mag_2}. Also, while metal-insulator transitions were observed in Eu$_2$Ir$_2$O$_7$ and Nd$_2$Ir$_2$O$_7$ \cite{Iridate_Exp_2, Iridate_Exp_3} and transport and optical behaviors were roughly consistent with a semimetal or a narrow band-gap semiconductor\cite{Iridate_Exp_3, Iridate_Exp_4}, the results overall were inconclusive. At the same, ARPES measurements are lacking, possibly because it has been difficult to grow large, high-quality single crystals or because it has been difficult to prepare a flat sample surface for measurement.

Shortly following the proposal for a time-reversal breaking Weyl semimetal in $R_2$Ir$_2$O$_7$, Burkov and Balents proposed, also in 2011, an engineered time-reversal breaking Weyl semimetal in a heterostructure built up from topological insulator and magnetic layers \cite{Burkov2011}. They provided a simple tight-binding model which explicitly includes the necessary ingredients for a Weyl semimetal and shows only two Weyl points, providing the ``hydrogen atom'' of a Weyl semimetal in theory. They also point out that their model shows a non-quantized anomalous Hall conductivity proportional to the separation of Weyl points in momentum space. Despite its theoretical significance, this model has remained unrealized in experiment. It requires a topological insulator and a trivial insulator with suitable lattice match that can both be grown by a thin film technique such as molecular beam epitaxy. In addition, magnetism must somehow be incorporated into the system, either by doping or perhaps by replacing the trivial insulator with a ferromagnet. Studying the emergent band structure of such a system in experiment would also be a formidable challenge.

Similar in spirit was a subsequent work by Bulmash \textit{et al.}, who in 2014 proposed engineering a time-reversal breaking Weyl semimetal by doping a topological insulator \cite{HgCdTe}. Specifically, they proposed starting with HgTe, a three-dimensional topological insulator, tuning it to the critical point for a band inversion by Cd doping and then introducing a magnetic order through Mn doping, giving Hg$_{1-x-y}$Cd$_x$Mn$_{y}$Te. Again, this proposal was quite reasonable in that it introduced the key ingredients for a Weyl semimetal in an explicit way through doping. Like the proposal by Burkov and Balents, the resulting system would also have only two Weyl points. However, despite the conceptual elegance of this work, ARPES measurements on Hg$_{1-x-y}$Cd$_x$Mn$_{y}$Te were unsuccessful. One reason might be that disorder from doping degraded sample quality too severely.

With the arrival of topological Dirac semimetals in Na$_3$Bi and Cd$_3$As$_2$ \cite{Chen_Na3Bi, Hasan2_Na3Bi, Hasan_Na3Bi, Hasan_Cd3As2, Chen_Cd3As2, Borisenko_Cd3As2}, the community also considered magnetic doping of a Dirac semimetal as a way to realize a time-reversal breaking Weyl semimetal, although there are no formal published proposals. These ideas were close to the work by Bulmash \textit{et al.}, but now the Dirac semimetal state was achieved intrinsically and only the time-reversal breaking needed to be implemented by doping. Indeed, these proposals faced difficulties similar to the case of Hg$_{1-x-y}$Cd$_x$Mn$_{y}$Te. Specifically, it was challenging to grow high-quality single crystals with magnetic doping which gave rise to a useful magnetic order. Another concern is that the spin-splitting from a magnetic doping may be too small to produce Weyl points above available experimental resolution and spectral linewidth in an ARPES measurement.

It was gradually appreciated that an intrinsic material may be easier to study than an engineered or doped system. However, finding an intrinsic time-reversal breaking Weyl semimetal presents its own challenges. The original proposal for $R_2$Ir$_2$O$_7$ is a case in point, because the magnetic order is difficult to predict from calculation and has not been conclusively shown in experiment. Also, it is unclear whether the value of the correlation parameter $U$ appropriate for the real material actually places the system in the Weyl semimetal phase. In 2011, the same year of the proposal for $R_2$Ir$_2$O$_7$ and the topological insulator multilayer, Xu \textit{et al}. proposed a Weyl semimetal in HgCr$_2$Se$_4$ \cite{HgCrSe}. Unlike $R_2$Ir$_2$O$_7$, HgCr$_2$Se$_4$ is known to be an intrinsic ferromagnet with a Curie temperature, $T_{\textrm{C}}\simeq120$ K, which is easily accessible in experiment. ARPES experiments on this system were not successful, perhaps because crystals are not of sufficiently high quality. However, recently, evidence for half-metallicity was reported in transport experiments, which may lead to renewed interest in this material \cite{HgCrSe_Transport}. Another interesting property is that the bands responsible for the band inversion arise from Hg and Se, while the ferromagnetism is associated with Cr. This makes the band inversion in \textit{ab initio} calculation robust to changes in $U$. Nonetheless, one concern is that the cubic structure of HgCr$_2$Se$_4$ allows many magnetization axes, which may favor the formation of small magnetic domains. It remains a considerable experimental challenge to create large magnetic domains \textit{in situ}, as well as to understand whether a Weyl semimetal could be shown in an ARPES experiment which averages over many magnetic domains. Nonetheless, the proposal for HgCr$_2$Se$_4$ offers an approach to searching for new candidates. In particular, it may be fruitful to study systems with an experimentally well-known magnetic order and where the Weyl semimetal state is robust to free parameters in the \textit{ab initio} calculation.

As the experimental complications of time-reversal breaking Weyl semimetals were appreciated, it was understood that breaking inversion symmetry may provide an easier route to the first Weyl semimetal. In retrospect, this should be rather clear. While the effect of breaking inversion symmetry and time-reversal symmetry is mathematically similar at the level of a single-particle Hamiltonian, these two symmetries relate to deeply different phenomena in any real material. Inversion symmetry breaking is a property of a crystal structure, which can be measured directly by X-ray diffraction and presents no particular complications to \textit{ab initio} calculation. By contrast, magnetism is a correlated phenomenon which is extremely difficult to reliably predict from first principles, challenging to understand in experiment and difficult to accurately capture in an \textit{ab initio} calculation even if the magnetic order is experimentally known. Additional complications arise because experiments must be carried out below the magnetic transition temperature and a measurement may average over many small magnetic domains. By contrast, large inversion breaking domains have been observed in inversion breaking materials. A well known example is the bulk Rashba material BiTeI, where the bulk Rashba splitting due to inversion breaking can be directly observed in ARPES. The idea that inversion breaking systems are simpler reignited the search for Weyl semimetals and led to the prediction and experimental observation of the first Weyl semimetal in TaAs.

Before discussing the theoretical prediction of TaAs, we note that despite the early focus on time-reversal breaking Weyl semimetals, there was some interest in inversion breaking systems. In 2012, Singh \textit{et al.} considered the tunable topological insulators TlBi(S$_{1-x}$Se$_{x}$)$_2$ and TlBi(S$_{1-x}$Te$_{x}$)$_2$ and proposed to engineer a Weyl semimetal by alternating between Se and Te from layer to layer, which would break inversion symmetry \cite{TlBiSe}. While this proposal was directly motivated by the recent experimental success in realizing a topological phase transition in TlBi(S$_{1-x}$Se$_{x}$)$_2$ \cite{Xu_TPT,TlBiSe,Hasan_Na3Bi}, the expected separation of the Weyl points in the proposed system fell below available experimental resolution in ARPES. Hal\'asz and Balents proposed in 2012 an inversion-breaking Weyl semimetal in a HgTe/CdTe heterostructure, inspired by a superlattice model similar to the time-reversal breaking model in a topological insulator multilayer \cite{Balents_HgTe}. This proposal remains unrealized, perhaps for reasons similar to the earlier proposal by Burkov and Balents, that there are considerable experimental challenges to studying the emergent band structure of such a superlattice. Lastly, in 2014, Liu and Vanderbilt proposed realizing an inversion breaking Weyl semimetal by tuning an inversion breaking topological insulator through a topological phase transition, exactly as first discussed by Murakami \cite{Vanderbilt}. Specifically, they predict a Weyl semimetal in LaBi$_{1-x}$Sb$_x$Te$_3$ and LuBi$_{1-x}$Sb$_x$Te$_3$ for $x \sim 38.5\%-41.9\%$ and $x \sim 40.5\%-45.1\%$. This proposal also remains unrealized, perhaps because the calculation results suggest that the Weyl semimetal phase is extremely sensitive to the composition $x$. Nonetheless, such tunable inversion breaking systems are of considerable interest because they would exhibit a topological phase transition to a Weyl semimetal, which to date is unrealized. Realizing such a system may require us to understand why the Weyl semimetal phase is so narrow in LaBi$_{1-x}$Sb$_x$Te$_3$ and LuBi$_{1-x}$Sb$_x$Te$_3$ and whether this behavior is natural.

The material search for intrinsic inversion breaking Weyl semimetals was based on a broad foundation of known crystal structures. These have been accumulated by X-ray diffraction experiments performed over the course of a century or more of research in physics and chemistry and which have been cataloged in databases such as the Inorganic Crystal Structure Database of FIZ Karlsruhe. Certain catalogs of magnetic compounds also exist with measurements of magnetic transition temperatures or magnetic structures. These sources provided a starting point for material searches. It is worth noting that nature prefers to maintain both inversion and time-reversal symmetry, severely limiting from the outset the number of possible material candidates for both kinds of Weyl semimetals.

Although \textit{ab initio} calculations for inversion-breaking systems are simpler, the calculation is still challenging because Weyl points typically occur at arbitrary $k$ points rather than at high symmetry points or on high symmetry lines. As a result, for each compound, it is necessary to calculate the band structure at all $k$ points throughout the bulk Brillouin zone to demonstrate or exclude the existence of Weyl points. A large spin-orbit coupling is preferred to produce a large spin-splitting. A closely-related requirement is that the Weyl points should be well-separated. Materials which also crystallize in an inversion-symmetric structure are excluded. The Weyl points should be very near the Fermi level and, for future transport experiments, it is preferable to find systems without irrelevant pockets at the Fermi level. The materials conditions such as cleavability and large single domain crystallinity that make ARPES measurements feasible were also considered to narrow down the candidate list further. In 2014, TaAs emerged as a promising candidate satisfying these ARPES criteria \cite{Weng2015, Huang2015}. Shortly thereafter, in early 2015, the observation of bulk Weyl fermions and topological Fermi arcs were finally reported in TaAs, demonstrating the first Weyl fermion semimetal \cite{Hasan_TaAs, TaAs_Ding}.

With the recipe for an inversion-breaking Weyl semimetal better understood, many other inversion-breaking Weyl semimetals were proposed \cite{SrSi2, LaAlGe, LaAlGe_2, WT-Weyl, WMoTe-Weyl, MT-Weyl, TaS, TIT, WP2, BiX}. Here, we do not attempt to survey these later inversion-breaking Weyl semimetal candidates. These systems will no doubt lead to many fascinating directions of research in the future.	
\bigskip
\bigskip
\bigskip
\section{\textbf{EXPERIMENTAL REALIZATION OF WEYL SEMIMETALS}}

Following independent theoretical predictions by the Princeton \cite{Huang2015} and Institute of Physics, Chinese Academy of Sciences (IOP) groups \cite{Weng2015}, the discovery of the first Weyl semimetal in TaAs followed quickly also by Princeton \cite{Hasan_TaAs} and IOP \cite{TaAs_Ding} independently. Experimental methods to demonstrate a topological origin of Fermi arc were shown earlier by Xu \textit{et.al.,} \cite{Hasan_Na3Bi}. Both bulk Weyl cones and topological Fermi arc surface states were directly observed by ARPES in bulk TaAs single crystals, demonstrating a Weyl semimetal at the same time in the bulk and on the surface and in excellent agreement with calculation. \cite{Hasan_TaAs}. Equivalently, a nonzero Chern number was directly measured from the Fermi arcs. These observations confirmed the bulk-boundary correspondence between Weyl cones and topological Fermi arcs. The discovery of the first Weyl semimetal in TaAs not only opened a new chapter in the study of topological phases of matter, but also validated the shift of the community from time-reversal breaking candidates to inversion breaking candidates. The identification of TaAs was quickly followed by the discovery of Weyl semimetals in NbAs \cite{NbAs_Hasan} and TaP \cite{TaP_Hasan, TaP_Shi}, with additional work on NbP \cite{NbP_Hasan, NbP_Feng, NbP_Chen, NbP_Ando}. Below, we review the experimental discovery of Weyl semimetals by ARPES in TaAs and its cousins. We also discuss general criteria for demonstrating a Weyl semimetal in ARPES, which arose from studies of these compounds \cite{NbP_Hasan}.

\subsection{Inversion symmetry breaking Weyl semimetal in TaAs}

\begin{figure*}
\centering
\includegraphics[width=16cm]{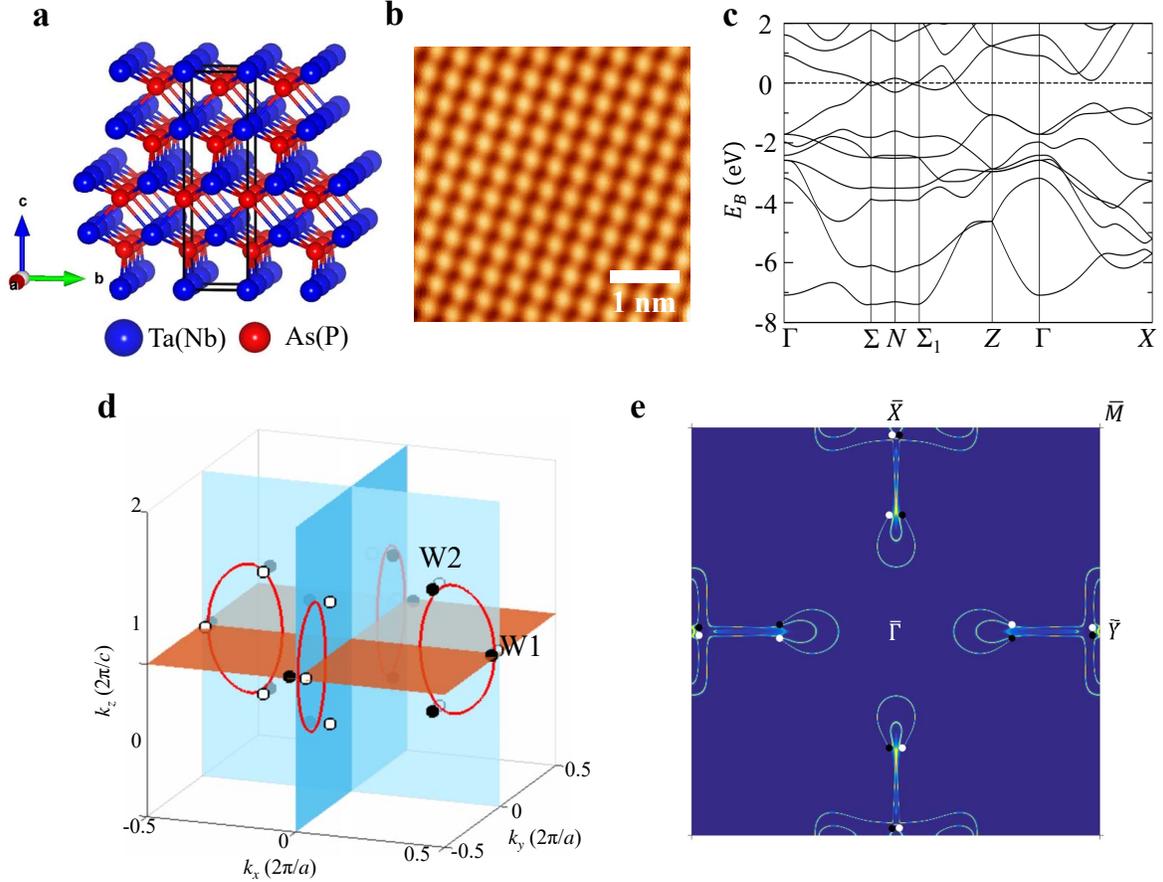}
\caption{\label{Weyl_THY}\textbf{The crystal and electronic structures of the Weyl semimetal TaAs.} \textbf{a,} Body-centered tetragonal structure of TaAs. The crystal lattice lacks an inversion center. \textbf{b,} Scanning tunneling microscopy (STM) topographic image of the (001) surface of TaAs, revealing a square lattice. \textbf{c,} Survey of the band structure of TaAs. At this level, the system is quite simple, with only two bands, a conduction and valence band, in the vicinity of the Fermi level, which approach each other on the $\Sigma-N-\Sigma_1$ line. \textbf{d,} The conduction and valence bands cross each other, forming nodal crossing points in $k$-space where the bulk energy gap vanishes. This panel shows the nodal crossing points in the bulk Brillouin zone. In the absence of spin-orbit coupling, the nodal crossing points are nodal-lines, i.e. 1D rings, on the $k_{x}=0$ mirror plane, $M_x$, and two nodal-lines on the $k_{y}=0$ mirror plane, $M_y$. In the presence of spin-orbit coupling, each nodal-line vaporizes into six 0D nodal points, the Weyl nodes. The Weyl nodes are denoted by small circles. Black and white show the opposite chiral charges of the Weyl points. We denote the 8 Weyl nodes located on the $k_z=2\pi/c$ plane as $W_1$ and the other 16 nodes away from this plane as $W_2$. \textbf{e,} Theoretically calculated (001) surface-state Fermi surface. Adapted with modifications from Ref. \cite{Hasan_TaAs}.}
\end{figure*}

TaAs crystallizes in a body-centered tetragonal lattice system. The lattice constants are $a=3.437$ $\textrm{\AA}$ and $c=11.656$ $\textrm{\AA}$, and the space group is $I4_1md$ ($\#109$). The crystal consists of interpenetrating Ta and As sub-lattices, where the two sub-lattices are shifted by $(a/2, a/2, \delta), \delta\simeq c/12$. Figure~\ref{Weyl_THY}\textbf{a} shows a schematic illustration of TaAs's crystal lattice. Crucially, the lattice breaks inversion symmetry, allowing it to realize a Weyl semimetal. The crystal can be viewed as a stack of square lattice layers of Ta or As (Fig.~\ref{Weyl_THY}\textbf{b}). A survey of the band structure shows that TaAs is a semimetal (Fig.~\ref{Weyl_THY}\textbf{c}). Further first-principles band structure calculations show that TaAs is a Weyl semimetal\cite{Weng2015,Huang2015}. Specifically, in the presence of spin-orbit coupling, the bands are non-degenerate everywhere in $k$ space except at the Kramers' points due to the breaking of inversion symmetry. These non-degenerate bands intersect at 24 discrete points, the Weyl points, in the bulk Brillouin zone (Fig.~\ref{Weyl_THY}\textbf{d}). On the (001) surface of TaAs, the 24 Weyl nodes project onto 16 points in the surface Brillouin zone. Eight projected Weyl points near the surface Brillouin Zone boundary ($\bar{X}$ and $\bar{Y}$ points) have a projected chiral charge of $\pm1$. The other eight projected Weyl points, close to midpoints of the $\bar{\Gamma}-\bar{X}(\bar{Y})$ lines, have a projected chiral charge of $\pm2$. Figure~\ref{Weyl_THY}\textbf{e} shows the calculated surface state Fermi surface. Near the midpoint of each $\bar{\Gamma}-\bar{X}(\bar{Y})$ line, there is a crescent-shaped feature, which consists of two curves that join each other at the two end points. The two curves of the crescent are two Fermi arcs, and the two end points correspond to the projected Weyl points with projected chiral charge $\pm2$.

\begin{figure*}
\centering
\includegraphics[width=16cm]{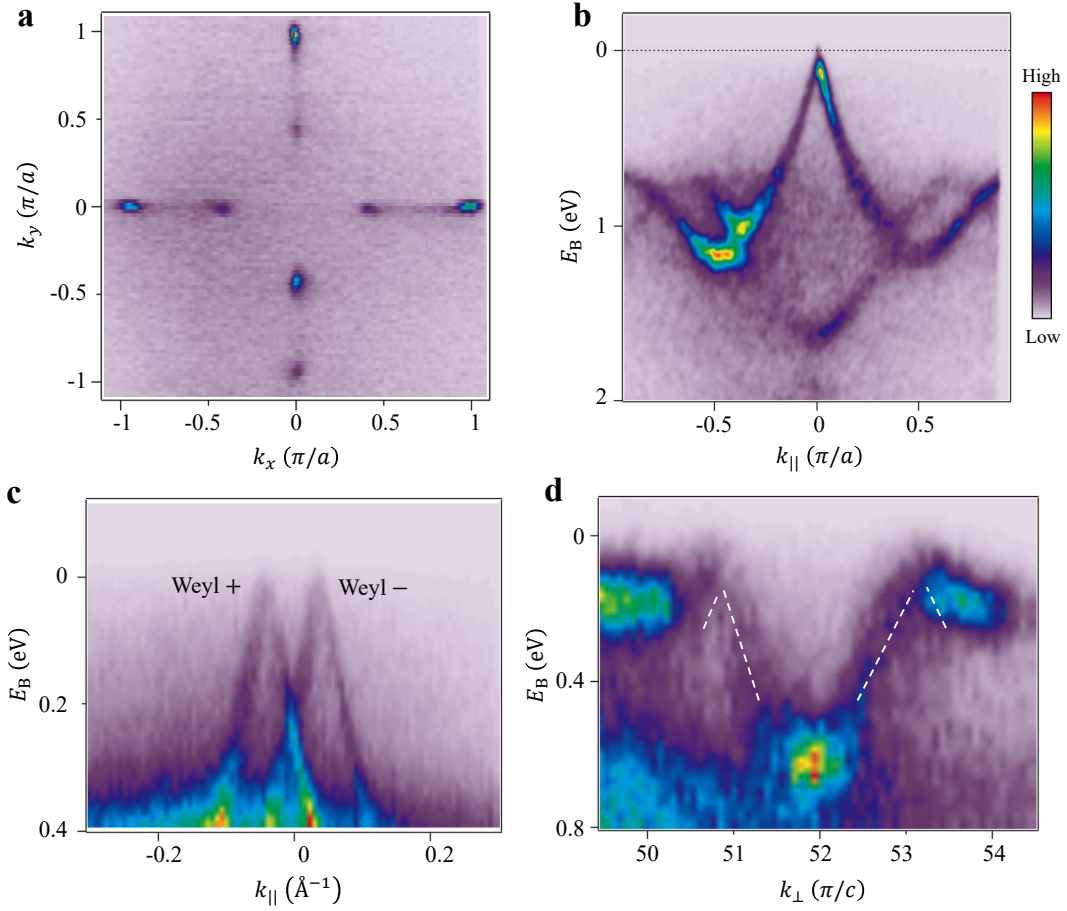}
\caption{\label{Weyl_Bulk}\textbf{Weyl fermions in TaAs.} \textbf{a,} ARPES-measured $k_x,k_y$ bulk Fermi surface of TaAs. The Fermi surface consists of discrete points that arise from the Weyl nodes. \textbf{b,} In-plane energy dispersion ($E_\textrm{B}-k_{\|}$) that goes through a $W_1$ Weyl node. A linear dispersion is clearly observed, consistent with the Weyl fermion cone. \textbf{c,} In-plane energy dispersion ($E_\textrm{B}-k_{\|}$) that goes through a pair of $W_2$ Weyl nodes. \textbf{d,} Out-of-plane energy dispersion ($E_\textrm{B}-k_{\perp}$) that goes through two $W_2$ Weyl nodes with the same $k_x,k_y$ but different $k_z$ value. Adapted  with modifications from Ref. \cite{Hasan_TaAs}. }
\end{figure*}

Both Weyl cones and Fermi arcs were observed by ARPES, each of which independently demonstrate the Weyl semimetal state in TaAs\cite{Hasan_TaAs, TaAs_Ding, TaAs_Ding_2}. The Weyl points and Weyl cones in TaAs were accessed by ARPES measurements at soft X-ray photon energies, which are sensitive to the bulk states. Soft X-ray data showed that the Fermi surface of TaAs consists of discrete points at specific incident photon energies and binding energies, at generic points in the Brillouin zone (Fig.~\ref{Weyl_Bulk}\textbf{a}). Away from these discrete points in binding energy, the bands dispersed linearly in both the in-plane momenta, $k_x$ and $k_y$ (Figs.~\ref{Weyl_Bulk}\textbf{b,c}) and the out-of-plane momentum $k_z$ (Figs.~\ref{Weyl_Bulk}\textbf{d}). The observation of linearly-dispersing band crossings at generic points in the bulk Brillouin zone demonstrated Weyl points and Weyl nodes in TaAs, showing that TaAs is a Weyl semimetal.

Independently of the bulk-sensitive soft X-ray ARPES measurements, surface-sensitive ARPES measurements at vacuum ultraviolet photon energies showed Fermi arcs on the (001) surface of TaAs. The ARPES measured Fermi surface showed three dominant features: a bowtie-shaped feature centered at the $\bar{X}$ point, a elliptical feature centered at the $\bar{Y}$ point, and a crescent-shaped feature near the midpoint of each $\bar{\Gamma}-\bar{X}(\bar{Y})$ line (Fig.~\ref{Weyl_Surf}\textbf{a}). The results were in excellent agreement with the \textit{ab initio} calculation (Fig.~\ref{Weyl_THY}\textbf{e}). More extensive APRES measurements of the crescent showed that it consists of two segments joined together at their endpoints (Fig.~\ref{Weyl_Surf}\textbf{b}), strongly suggesting Fermi arcs. \textit{Ab initio} calculation showed topological Fermi arcs in close agreement with the crescent observed in ARPES, showing that Fermi arcs were observed. A non-zero Chern number was also directly observed, providing another demonstrating of Fermi arcs from the ARPES data, see discussion below. Finally, ARPES data also showed that the terminations of the Fermi arcs coincide with the projections of the Weyl nodes, which demonstrated the bulk-boundary correspondence principle (Fig.~\ref{Weyl_Surf}\textbf{d}), further confirming the observation of a Weyl semimetal in TaAs \cite{Hasan_TaAs}. Later measurements by spin-resolved ARPES showed the spin texture of the Fermi arcs \cite{Hasan_TaAs_spin, Ding_TaAs_spin}. In calculation, the spin texture was found to wind against the dispersion of the Fermi arc, so that if we traverse the Fermi arc in a clockwise way, the spin texture winds in a counter-clockwise way (Fig.~\ref{Weyl_Surf}\textbf{c}). This spin texture was directly measured by spin-resolved ARPES and was found to be in excellent agreement with the \textit{ab initio} calculation (Fig.~\ref{Weyl_Surf}\textbf{e,f}). The relationship of the spin texture to the topological invariants of the bulk is not understood. It appears that there is no obvious constraint placed on the Fermi arc spin texture by the chiral charge of the Weyl points. We note that no spin-resolved soft X-ray ARPES end-station is currently available, which has hindered the measurement of the spin texture of the Weyl points and Weyl cones. This has prevented the direct experimental measurement of the chiral charge in a Weyl semimetal. To directly measure the chiral charge, either the appropriate ARPES system must become available or a new material is needed where the bulk band structure can be accessed by vacuum ultraviolet ARPES.

A non-zero Chern number in TaAs was directly measured from an ARPES spectrum of the surface states \cite{NbP_Hasan}. The authors considered a cylindrical tube in the bulk Brillouin zone enclosing two Weyl points of the same chiral charge. The Chern number on this two dimensional manifold is $+2$ (Fig.~\ref{Weyl_Loop}\textbf{a}). As a result, the one-dimensional edge of the cylindrical slice hosts protected chiral edge states which cross the bulk band gap and have net chirality $+2$ (Fig.~\ref{Weyl_Loop}\textbf{b}). In this way, a Chern number can be directly measured in an ARPES spectrum of the surface state band structure by counting the net chirality of surface state crossings on a closed loop in momentum space (Fig.~\ref{Weyl_Loop}\textbf{c}). It was found in ARPES that the one-dimensional band structure on the loop showed two edge states of the same chirality at the Fermi level, in agreement with the topological theory (Fig.~\ref{Weyl_Loop}\textbf{d}). The direct measurement of a non-zero Chern number in TaAs by ARPES provides yet another independent demonstration that TaAs is a Weyl semimetal.

\subsection{Criteria for topological Fermi arcs in Weyl semimetals}

The case of TaAs showed that topological Fermi arcs do not always appear as disjoint arcs. Specifically, in TaAs, the Fermi arcs appeared in pairs which together formed a closed contour. As a result, the signature of a Fermi arc in the experiments on TaAs was not a disjoint arc but rather a Chern number or a surface state kink, as discussed above. More generally, it was directly relevant for experiment to understand the ways that a topological Fermi arc can arise in a material. As far as is currently understood, there are four distinct criteria for topological Fermi arcs \cite{NbP_Hasan}. All signatures are in principle experimentally accessible in an ARPES measurement of a surface state band structure. Each signature alone, observed in any set of surface state bands, is sufficient to show that a material is a Weyl semimetal,

\begin{enumerate}
\item \label{dis} \textit{Disjoint arc}: Any surface state constant-energy contour with an open curve is a Fermi arc and demonstrates a Weyl semimetal.
\item \label{kink} \textit{Kink off a rotation axis}: A Weyl point is characterized by chiral charge $n$, equal to the Chern number on a small spherical manifold enclosing the Weyl point in the bulk Brillouin zone, illustrated by the small sphere in Fig. \ref{Weyl_Loop}(e) \cite{Wan2011, WeylTransportReviewQi, TI_book_2014_2}. For a Weyl point of chiral charge $|n| > 1$ or if multiple Weyl points project onto the same point of the surface Brillouin zone, there may not be a disjoint constant-energy contour because multiple arcs will emanate from the same Weyl point projection. However, the arcs will generically attach to the Weyl point at different slopes, giving rise to a kink in the constant-energy contour. Moreover, such a kink can only arise from the attachment of two Fermi arcs. A kink on the projection of a rotation axis may arise in a topological Dirac semimetal \cite{Hasan_Na3Bi}. Off a rotation axis, such a kink guarantees a Weyl semimetal.
\item \label{odd} \textit{Odd number of curves}: For projected chiral charge $|n| > 1$, the constant-energy contours may consist entirely of closed curves and the kink may be below experimental resolution, so the constant-energy contour appears everywhere smooth. However, if $|n|$ is odd, the constant-energy contour will consist of an odd number of curves, so at least one curve must be a Fermi arc.
\item \label{Chern} \textit{Non-zero Chern number}: Consider any closed loop in the surface Brillouin zone where the bulk band structure is everywhere gapped and, at some energy, add up the signs of the Fermi velocities of all surface states along this curve, with $+1$ for right movers and $-1$ for left movers. The sum is the projected chiral charge enclosed in the curve, corresponding to a Chern number on a bulk \cite{Wan2011, WeylTransportReviewQi, TI_book_2014_2}. A non-zero sum on at least one loop demonstrates a Weyl semimetal, provided the loop is chosen to be contractible on the torus formed by the surface Brillouin zone.
\end{enumerate}

\noindent Note that while (\ref{dis}), (\ref{kink}) and (\ref{odd}) describe properties of a constant-energy slice of the Fermi surface, the counting argument (\ref{Chern}) requires a measurement of the dispersion. We note also that criterion (\ref{Chern}) allows us to determine all bulk topological invariants and Weyl points of a material by studying only its surface states.

One caveat in this rather formal analysis is that in a real ARPES experiment we never rigorously measure only the surface or bulk states, but rather some combination of the two. There are, furthermore, many experimental scenarios where a surface state unrelated to Fermi arcs can na\"ively satisfy one of the criteria. For instance, the photoemission cross section of certain bands or certain regions of a band may be suppressed under particular measurement conditions. This effect can give rise to an apparently disjoint contour or an apparent Chern number in a completely topologically trivial material. In addition, a kink will always be smeared out in an experiment, so that it can be challenging to distinguish between a kink and a smooth but rapidly dispersing band. Therefore, an \textit{ab initio} calculation remains crucial for ARPES studies of new Weyl semimetal candidates. However, it is equally unacceptable to simply show a general agreement between an ARPES spectrum and a calculation. We note that a number of early works claimed to show a Weyl semimetal by ARPES in TaAs and NbP by measuring a Chern number, but those measurements depended on resolving Fermi arcs which clearly fell below any available experimental resolution or spectral linewidth \cite{TaAs_Ding, TaAs_Chen, NbP_Feng, NbP_Ando, NbP_Chen}. As a result, those works essentially used a rough, overall agreement between calculation and experiment to claim a Weyl semimetal. A reasonable standard for detection of a Weyl semimetal in ARPES is to consider a set of surface states which can be confirmed in calculation to be topological Fermi arcs and which can be clearly identified in an ARPES spectrum. These Fermi arcs will not always show up as disjoint arcs, but they must satisfy at least one of the four criteria discussed here.

\subsection{Weyl fermion transport and signatures of the chiral anomaly}

It is known in quantum field theory that quantum fluctuations can violate classical conservation laws, a phenomenon called a quantum anomaly \cite{Anomaly}. Perhaps the best-studied example is the chiral anomaly associated with Weyl fermions \cite{nielsen1983adler, ABJ1, ABJ2}. Historically, the chiral anomaly was crucial in understanding a number of important aspects of the Standard Model of particle physics, such as the triangle anomaly associated with the decay of the neutral pion $\pi^{0}$, Refs. \cite{ABJ1, ABJ2}. Despite having been discovered more than 40 years ago, quantum anomalies remained solely in the realm of high-energy physics.

The discovery of Weyl fermion semimetals provides a natural route to realizing the chiral anomaly in condensed matter physics. Consider a Weyl semimetal with a particular configuration of Weyl points of positive and negative chiral charge in the bulk Brillouin zone. Parallel magnetic and electric fields can pump electrons between Weyl cones of opposite chirality, giving rise to a population imbalance between Weyl cones of positive and negative chiral charge. This means that the numbers of Weyl fermions of a given chirality are not separately conserved \cite{nielsen1983adler, Duval06, Fukushima, Son13, Burkov14}. The key observable consequence of the chiral anomaly in a condensed matter system is that the longitudinal resistance is predicted to decrease as a function of an external magnetic field, giving rise to a negative longitudinal magnetoresistance (LMR). Except for this natural platform, Dirac semimetals, a class of materials that host Dirac fermion quasiparticles, may similarly give rise to a negative LMR under external magnetic field. In that case, the magnetic field not only directly produces a chiral anomaly but additionally serves the purpose of splitting the Dirac cone into a pair of Weyl cones of opposite chirality by breaking time-reversal symmetry. It is worth noting that under such conditions, extra caution is needed, because in real materials a magnetic field may have many effects other than a simple Zeeman splitting of the band structure \cite{suzuki}.

The negative LMR was directly detected in electrical transport experiments on the TaAs family\cite{Chiral_anomaly_Jia, Chiral_anomaly_ChenGF, NbAs_Transport_1, NbAs_Transport_2, NbAs_Transport_3, NbAs_Chiral_anomaly_1, TaP_Transport_1, TaP_Transport_2, Du J, Arnold F} (Figs.~\ref{Weyl_Chiral}\textbf{a-c}) and a number of Dirac semimetals \cite{Korean_BiSb,Ong_Chiral,CdAs_Chiral}. Other supporting evidence includes: (1) The negative LMR was prominent only in the geometry of parallel electric and magnetic fields. This is consistent with the $\vec{E}\cdot\vec{B}$ term in the chiral anomaly formulation (Fig.~\ref{Weyl_Chiral}\textbf{c}). (2) The negative LMR did not depend on the electrical current direction with respect to the crystalline axis (Figs.~\ref{Weyl_Chiral}\textbf{a,b}). However, these data are not conclusive. A number of other effects can also lead to a negative LMR in metals\cite{GMR,Pippard,Ag2Se,Argyres,InSb, Kikugawa,Das_Sarma_chiral}, and some of them\cite{Argyres,Das_Sarma_chiral} may show the same systematic dependences as described above. Hence, to achieve an unambiguous proof, further study is needed. One particular phenomenon that may provide stronger evidence is the dependence of the LMR on chemical potential. Because Weyl points are monopoles of Berry curvature, the magnitude of negative LMR is expected to follow a dramatic $1/E_\textrm{F}^2$ dependence as the Fermi level moves away from the energy of the Weyl points. This $E_\textrm{F}$ dependence\cite{Chiral_anomaly_Jia} (Fig.~\ref{Weyl_Chiral}\textbf{d}) can distinguish the chiral anomaly from other negative LMR effects and, therefore, provides a clearer demonstration of the chiral anomaly due to Weyl fermions.

\bigskip
\bigskip
\bigskip

\section{\textbf{OUTLOOK}}
It has been less than a year since the experimental realization of the first Weyl semimetals, and the field has already evolved dramatically. One topic of recent interest is the realization in a material of a strongly Lorentz-violating Weyl fermion. Traditionally, studies of Weyl fermions in quantum field theory were concerned with applications to particle physics, where Lorentz symmetry is respected. However, low-energy effective field theories in crystals need not satisfy Lorentz invariance, providing a richer variety of allowed theories. In particular, the form of the dispersion of a Weyl fermion in a crystal has more freedom than in particle physics. Recently, it was pointed out that this freedom allows for a novel type of Weyl fermion where the Weyl cone is tipped over on its side \cite{Lorentz, WT-Weyl, Bergholtz, Trescher}. Such strongly Lorentz-violating, or Type II, Weyl fermions allow us to study, in table-top experiments, exotic Lorentz-violating theories that are beyond the Standard Model. They also open up experimental opportunities for studying novel spectroscopic and transport phenomena specific to Type II Weyl fermions. Such phenomena include a chiral anomaly associated with a transport response that depends strongly on the direction of the electric field, an antichiral effect of the chiral Landau level, a modified anomalous Hall effect, and emergent Lorentz invariance arising from electron–electron interactions \cite{Lorentz, WT-Weyl, Bergholtz, Trescher, Beenakker, Zyuzin, Isobe}. To date, Type II Weyl fermions have only been suggested in W$_{1-x}$Mo$_x$Te$_2$ \cite{WT-Weyl, WMoTe-Weyl, MT-Weyl} and observed in LaAlGe \cite{LaAlGe, LaAlGe_2}. Therefore, it is of importance to continue the study of Type II Weyl semimetals.

Because the TaAs family exhibits twelve pairs of Weyl points, it is of some interest to find simpler materials with fewer Weyl points. Material searches are under way to find the ``hydrogen atom'' versions of Weyl semimetals with the minimum number of Weyl points possible, either four Weyl points in inversion-breaking Weyl semimetals or two Weyl points in magnetic Weyl semimetals. An additional challenge is to tune the materials so that the Fermi level is close to the Weyl points, ideally within $5$-$10$ meV, without irrelevant pockets near the Fermi level, so that the Weyl fermionic excitations constitute the dominant transport channel. Moreover, the $\mathcal{T}$-breaking Weyl semimetals and Weyl superconductors (both $\mathcal{T}$-breaking and $\mathcal{I}$-breaking) can give rise to a wide range of novel properties. To understand the interplay between the electronic interaction and the topological state in Weyl fermion semimetals, $\mathcal{T}$-breaking magnetic Weyl materials could be crucial because magnets already harbor strong interactions. From a purely mathematical point of view, interacting Weyl phases further broaden the classification of topological phases of matter. Exploring the nontrivial spin polarization properties of interacting Weyl materials could reveal a rich phase diagram. In a loose sense, the spin texture is approximately proportional to the Berry flux and, hence, projects like monopoles or antimonopoles near the Weyl nodes in momentum space. This opens up opportunities for applications in spintronics. Realizing Weyl superconductors would be another exciting frontier. The Majorana surface states of a Weyl superconductor can be potentially used for topological qubits. Given the rapid development of the field, it is also quite possible that in a few years, the most exciting frontier will be something not projected here.

\section{\textbf{ACKNOWLEDGMENTS}}

The authors thank Adam Kaminski, Arun Bansil, BaoKai Wang, Bingbing Tong, Cheng Guo, Chenglong Zhang, Chi-Cheng Lee, Chi Zhang, Chuang-Han Hsu, Daixiang Mou, Daniel S. Sanchez, Donghui Lu, Fangcheng Chou, Fumio Komori, Guang Bian, Guoqing Chang, Hai-Zhou Lu, Hao Zheng, Hong Lu, Horng-Tay Jeng, Hsin Lin, J. D. Denlinger, Jie Ma, Junfeng Wang, Kenta Kuroda, Koichiro Yaji, Lunan Huang, Madhab Neupane, Makoto Hashimoto, Mykhailo L. Prokopovych, Nan Yao, Nasser Alidoust, Pavel P. Shibayev, Raman Sankar, Shik Shin, Shuang Jia, Shun-Qing Shen, Sungkwan Mo, Takeshi Kondo, Tay-Rong Chang, Titus Neupert, Vladimir N. Strocov, Xiao Zhang, Yun Wu, Zhujun Yuan and Ziquan Lin for collaborations. Work at Princeton University by S.-Y.X. and M.Z.H. is supported by the US Department of Energy under Basic Energy Sciences Grant No. DOE/BES DE-FG-02-05ER46200 and Grant No. DE-AC02-05CH11231 at the Advanced Light Source at Lawrence Berkeley National Laboratory (LBNL), and Princeton University funds. M.Z.H. acknowledges Visiting Scientist user support from LBNL and partial support from the Gordon and Betty Moore Foundation under Grant No. GBMF4547/Hasan. Shin-Ming Huang acknowledges visiting scientist support from Princeton University.

\clearpage
\begin{figure*}
\centering
\includegraphics[width=16cm,trim={100 0 100 0},clip]{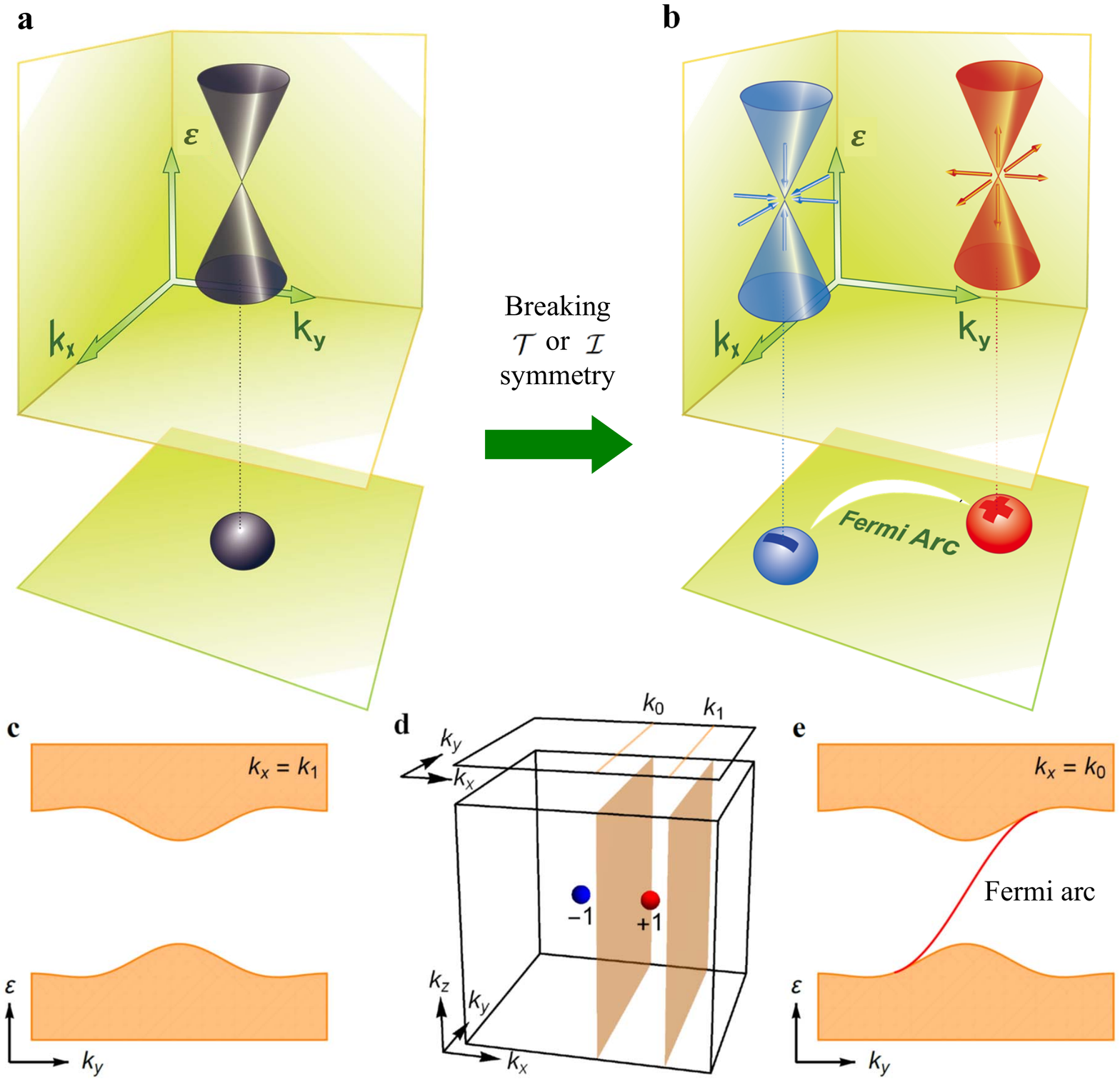}
\end{figure*}

\clearpage
\begin{figure*}
\centering
\caption{\label{Weyl_Cartoon}\textbf{Weyl fermions on a lattice.} \textbf{a,} One way to create a Weyl semimetal is to start with a massless Dirac fermion in a system with both time reversal and inversion symmetries. Such a Dirac fermion corresponds to the intersection of two doubly-degenerate bands and can be realized, for instance, at the critical point of a topological phase transition between a normal insulator and a topological insulator. \textbf{b,} By breaking time reversal or inversion symmetry, the Dirac fermion splits into a pair of Weyl fermions of opposite chiralities. Each Weyl fermion is a monopole or anti-monopole of Berry curvature and, equivalently, is associated with a Chern number. The Chern number guarantees the existence of a topological Fermi arc surface state that connects the projections of the Weyl points in the surface Brillouin zone. \textbf{c-e,} Weyl semimetals are characterized by Chern numbers, as in the integer quantum Hall effect. For instance, we can consider a system with two Weyl points of chirality $\pm1$ and we can calculate the Chern number on slices of the Brillouin zone at different $k_x$, \textbf{d}. When a slice is swept through a Weyl point, the two-dimensional system undergoes a topological phase transition and the Chern number changes by $\pm 1$. For slices with a Chern number $\nu = 0$, the one-dimensional edge of the two-dimensional slice is gapped, \textbf{c}, while slices with a Chern number $\nu = +1$ host a protected gapless chiral edge mode, \textbf{e}. The Fermi arc can be understood as arising from all of the chiral edge states assembled together into a surface state.}
\end{figure*}

\clearpage

\clearpage

\begin{figure*}
\centering
\includegraphics[width=16cm]{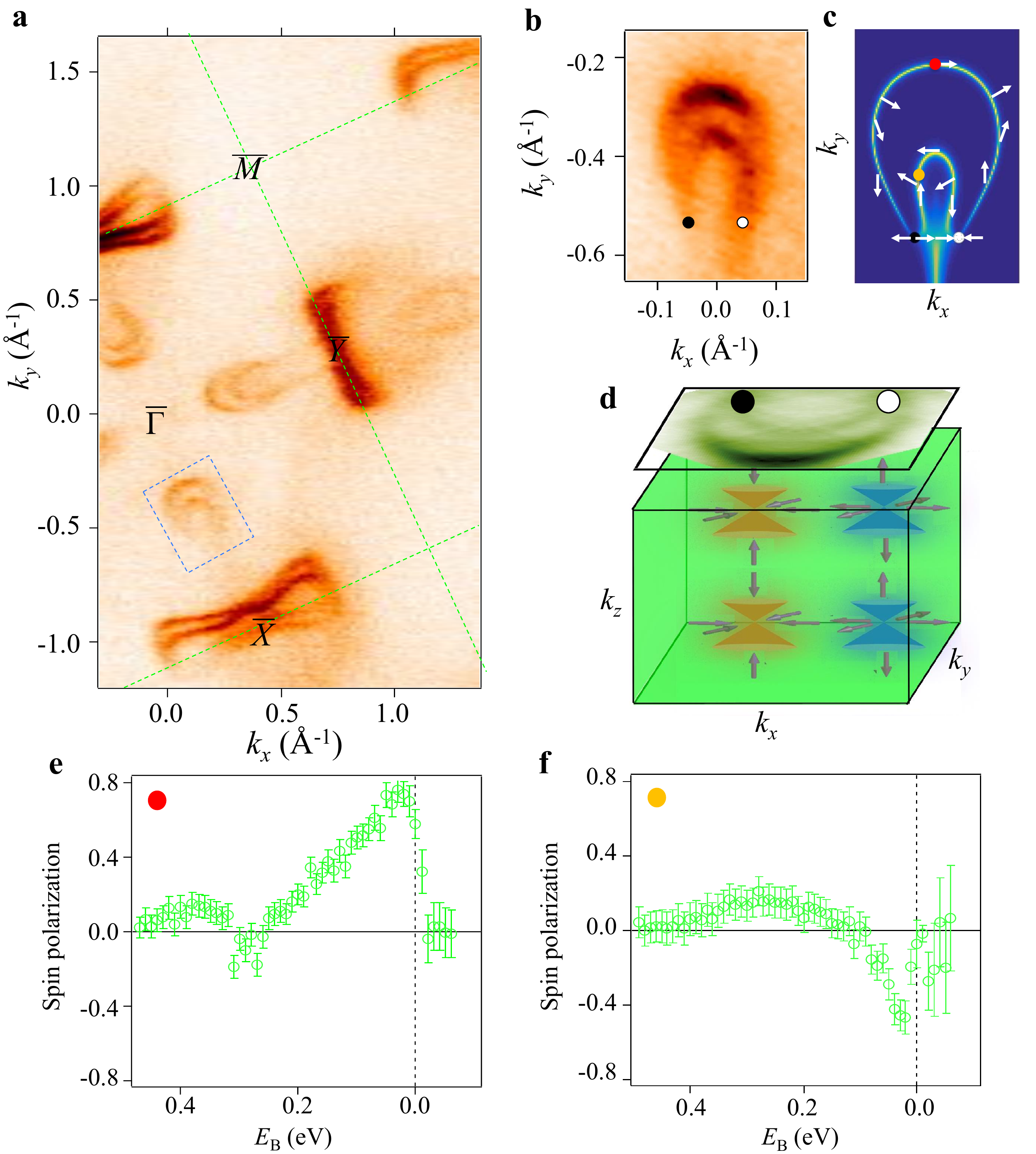}
\end{figure*}

\begin{figure*}
\centering
\caption{\label{Weyl_Surf}\textbf{Topological Fermi arcs in TaAs.} \textbf{a,} ARPES spectrum of the surface state band structure of TaAs near the Fermi level, $E_F$. The green dotted lines denote the boundaries of the surface Brillouin zone. \textbf{b,} High resolution ARPES Fermi surface of the double Fermi arcs that arise from a pair of projected $W_2$ Weyl nodes. The $k$-space region of this map is indicated by the blue box in panel (a). \textbf{c,} Theoretically calculated Fermi surface of the double Fermi arcs that arise from the same pair of projected $W_2$ Weyl nodes. We indicate schematically the spin texture of the Fermi arcs. \textbf{d,} Lower box: illustration of four W$_2$ Weyl nodes in the bulk Brillouin zone, two of each chirality. Upon projection on the (001) surface Brillouin zone, the two $+1$ and the two $-1$ Weyl nodes project onto each other, giving a projected chiral charge of $\pm2$. Top surface: ARPES spectrum of the two Fermi arcs connecting the projected Weyl nodes. The black and white circles in panels (b-d) show the projected $W_2$ Weyl nodes with opposite chiralities. \textbf{e,f,} Spin polarization along the $k_x$ direction, as measured in spin-resolved ARPES at two points on the Fermi arcs, as indicated by the red and orange dots in (c). Adapted  with modifications from Refs. \cite{Hasan_TaAs, Hasan_TaAs_spin}.}
\end{figure*}

\begin{figure*}
\centering
\includegraphics[width=14cm]{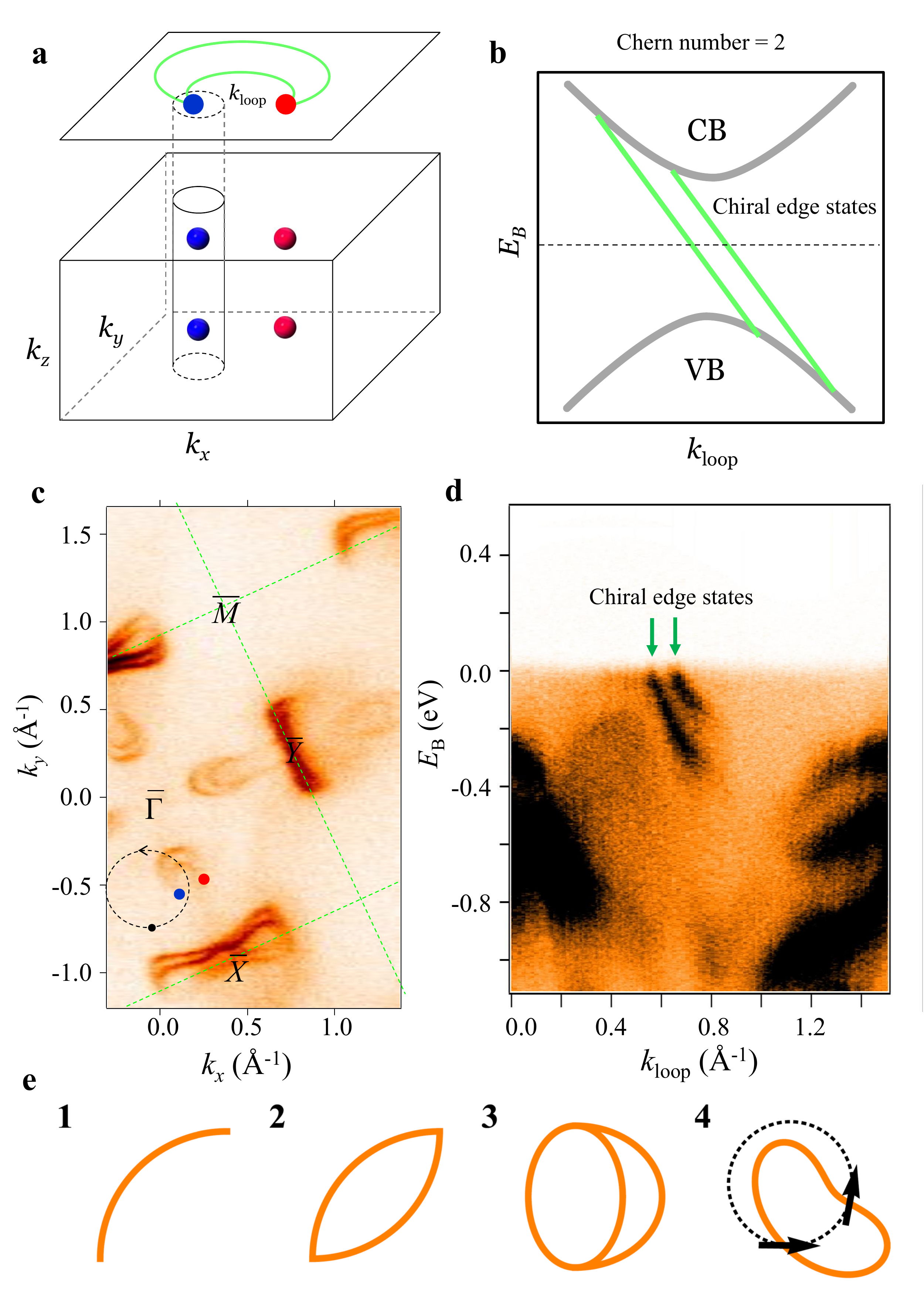}
\end{figure*}

\begin{figure*}
\centering
\caption{\label{Weyl_Loop}\textbf{Bulk-boundary correspondence in the Weyl semimetal TaAs.} \textbf{a,} Illustration of a Chern number in TaAs. The blue and red dots are Weyl points of opposite chiralities. We consider a cylindrical tube extending through the bulk Brillouin zone and enclosing two Weyl points of the same chirality. There is a net enclosed chiral charge of $+2$, so the Chern number on this two-dimensional slice of the Brillouin zone is $+2$. \textbf{b,} On the one-dimensional boundary of the tube, the Chern number protects two chiral edge states, which make up one slice of the topological Fermi arcs. \textbf{c,} We can directly demonstrate that TaAs is a Weyl semimetal from the surface state band structure as measured in ARPES by drawing a loop in the surface Brillouin zone and counting the crossings to show a nonzero Chern number. The loop is shown by the dotted black line. \textbf{d,} The surface states along the loop. We find two edge states of the same chirality, showing a Chern number of $+2$. Note that for a conventional electron or hole pocket, such a counting argument will always give $0$. (e) The four criteria for a topological Fermi arc. (1) A disjoint contour. (2) A closed contour with a kink. (3) No kinks within experimental resolution, but an odd set of closed contours. (4) An even number of contours without kinks, but net non-zero chiral edge modes. Adapted  with modifications from Refs. \cite{Hasan_TaAs, NbP_Hasan}.}
\end{figure*}

\clearpage

\begin{figure*}
\centering
\includegraphics[width=16cm,trim={40 80 40 80},clip]{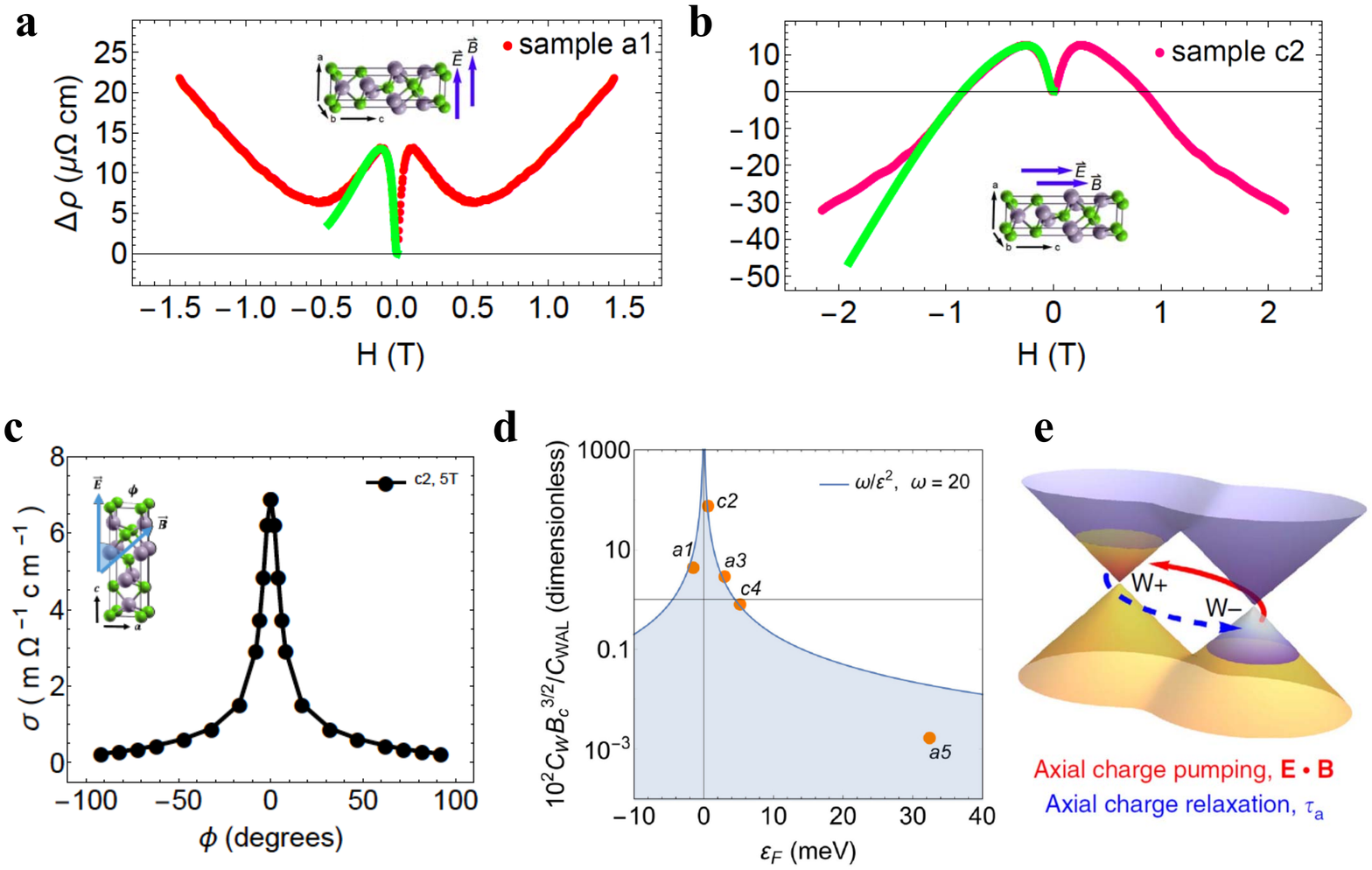}
\end{figure*}

\begin{figure*}
\centering
\caption{\label{Weyl_Chiral}\textbf{Signatures of the Adler-Bell-Jackiw chiral anomaly in TaAs.} \textbf{a,b,} Longitudinal magnetoresistance (LMR) at $T=2$ K, for two samples a1 and c2, Ref. \cite{Chiral_anomaly_Jia}. See again Ref. \cite{Chiral_anomaly_Jia} for complete data on additional samples a3, c4 and a5 discussed in that study. The green curves are fits to the LMR in the semiclassical regime. \textbf{c,} Magnetoconductivity as a function of the angle between the $\vec{E}$ and $\vec{B}$ fields. The $y$ axes of panels (a-c) are the change of the resistivity with respect to the zero-field resistivity, $\Delta\rho=\rho(B)-\rho(B=0)$, or the magnetoconductivity as determined from $\Delta \rho$. \textbf{d,} Dependence of the chiral coefficient $C_{\textrm{W}}$, appropriately normalized by the other fitting coefficients, on chemical potential, $E_{\textrm{F}}$. Remarkably, the observed scaling behavior is $1/E_{\textrm{F}}^2$, as expected from the dependence of the Berry curvature on chemical potential in the simplest model of a Weyl semimetal, $\Omega \propto 1/E_{\textrm{F}}^2$. \textbf{e,} A cartoon illustrating the chiral anomaly in TaAs\cite{Chiral_anomaly_Jia}. The chiral anomaly leads to an axial charge pumping, for $\vec{E}\cdot\vec{B} \neq 0$. This causes a population imbalance between Weyl cones of opposite chiralities. The charge pumping reaches an equilibrium with the axial charge relaxation, characterized by a timescale $\tau_a$, Refs. \cite{Son13, Burkov14}. Note that the axial charge relaxation time $\tau_a$ can be directly obtained from the observed negative LMR data through the chiral coefficient $C_{\textrm{W}}=e^4\tau_{\textrm{a}}/{4\pi^4\hbar^4g(E_{\textrm{F}})}$, as discussed elsewhere\cite{Chiral_anomaly_Jia}. Adapted with modifications from Ref. \cite{Chiral_anomaly_Jia, OurNatMatRev}.}
\end{figure*}

\clearpage

\end{document}